\documentclass[aps, prd, 10pt, twocolumn]{revtex4-1}

\usepackage{graphicx}
\usepackage{amssymb, amsmath,mathtools}
\usepackage{bbm}
\usepackage{wasysym}
\usepackage{slashed}

\graphicspath{{./}{./plots/}}

\DeclareMathOperator{\Tr}{Tr}

\DeclareMathOperator{\artanh}{artanh}

\newcommand{\Nchi}{N_{\mathrm{c}}^{\chi\mathrm{SB}}}
\newcommand{\Nconf}{N_{\mathrm{c}}^\mathrm{conf}}

\begin{document}

\title{Spontaneous breaking of Lorentz symmetry in $(2+\epsilon)$-dimensional QED}

\author{Lukas Janssen}
\email{lukas.janssen@tu-dresden.de}
\affiliation{Institut f\"ur Theoretische Physik, Technische Universit\"at Dresden, 01062 Dresden, Germany}

\begin{abstract}
The phase diagram of massless quantum electrodynamics in three space-time dimensions as a function of fermion flavor number $N$ exhibits two well-known phases: at large $N > \Nconf$ the system is in a conformal gapless state, while for small $N < \Nchi$ the fermions are expected to develop a dynamical mass due to spontaneous chiral symmetry breaking. Using $\epsilon$~expansion near the lower critical dimension of $2$, as well as the recent results on the generalization of the $F$ theorem to continuous dimension, we show that $\Nconf > \Nchi$. There is therefore an intermediate range of values of $N$ at which a third phase is stabilized. We demonstrate that this phase is characterized by spontaneous breaking of Lorentz symmetry, in which a composite vector boson field acquires a vacuum expectation value with the fermions and the photon remaining massless. 
\end{abstract}

\maketitle

\section{Introduction}

Many theories that attempt to unify gravity with quantum mechanics, such as string theory~\cite{kostelecky1989} and loop quantum gravity~\cite{gambini1999, sudarsky2002}, predict small violations of relativity. 
Currently, enormous efforts are made to test the validity of Lorentz symmetry at ultrahigh precision, ultimately aiming for an experimental access to Planck-scale physics~\cite{toma2012, disciacca2013, pruttivarasin2015, michimura2013}.
However, when coupled to gravity, explicit breaking of Lorentz symmetry is considered inconsistent with general arguments~\cite{kostelecky2004}. It is therefore usually expected that if at all any Lorentz symmetry breaking does happen, it can do so only \emph{spontaneously}~\cite{bluhm2006}. While various theories that exhibit spontaneous Lorentz symmetry breaking have been discussed in the past~\cite{kostelecky1989, andrianov1995, kostelecky2004, bluhm2006, bluhm2015}, a simple toy model in which the effect is not artificially ``put in by hand'' appears to be rare~\cite{hosotani1993}.

In this work, we argue that such a role may be played by an old acquaintance from a different corner of the field-theory literature: massless quantum electrodynamics in three space-time dimensions (QED$_3$). While initially conceived as a prototype model for strongly coupled and asymptotically free gauge theories~\cite{pisarski1984}, the theory has more recently provoked considerable renewed interest in the condensed matter community, initiated by the proposal of QED$_3$ as an effective model for high-temperature superconductors~\cite{rantner2001, franz2001, herbut2002}. Today, the theory and variants thereof are intensely debated as low-energy theories of U(1) Dirac spin liquids~\cite{hermele2004, ran2007, xu2008, he2015, wang2016} and of quantum critical points between symmetry-protected topological phases~\cite{grover2013, lu2014}, as field theory for the half-filled Landau level state~\cite{son2015}, and as dual descriptions of surface states of topological insulators~\cite{wang2015, metlitski2015, mross2015}.

QED$_3$ is believed to exhibit two distinct phases~\cite{appelquist1988}. If the number $N$ of massless fermion flavors is large, the theory has an interacting conformal ground state.  
At low $N$, on the other hand, the fermions are expected to acquire a dynamical mass via spontaneous chiral symmetry breaking. 
Obviously, knowledge of the true ground state of the theory at a given $N$ is indispensable for its interpretation within the condensed-matter applications.
Since the initial analysis in the 1980's~\cite{appelquist1988}, many follow-up works have attempted to establish the phase diagram. These include higher-order $1/N$ calculations~\cite{nash1989}, various nonperturbative approximations of the Dyson-Schwinger equations~\cite{maris1996, fischer2004, goecke2009}, different types of renormalization group (RG) calculations~\cite{kubota2001, kaveh2005, braun2014, dipietro2016}, and lattice Monte Carlo simulations~\cite{hands2002, raviv2014, karthik2016}. 
Lately, progress has been made by exploiting the generalization of the 
central charge theorem to three dimensions---the ``$F$ theorem''~\cite{grover2014}. A ``generalized $F$ theorem'' that may be valid for arbitrary integer and noninteger dimension has also been applied recently~\cite{giombi2015b, giombi2016}.
The common theme of all these previous works is the conjecture that the conformal large-$N$ state becomes unstable below some critical $\Nconf$, giving rise to a \emph{direct} transition towards a chiral-symmetry-broken state.

\begin{figure}[tb]
\includegraphics[scale=1.1]{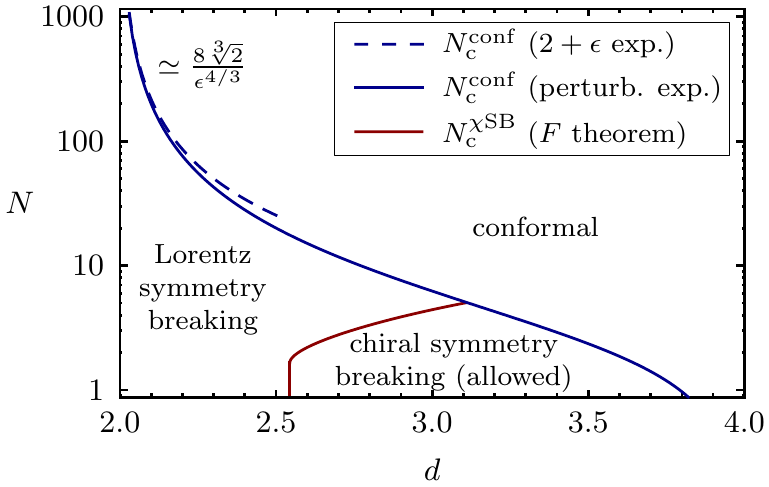}
\caption{Phase diagram of QED$_d$ as a function of space-time dimension $d$ and fermion flavor number $N$. Blue lines: $\Nconf$ from $2+\epsilon$ expansion (dashed) and perturbative expansion in fixed dimension (solid). Red line: Upper bound for $\Nchi$ from the $F$ theorem.}
\label{fig:phase-diagram}
\end{figure}

In this paper, we significantly correct this largely accepted scenario.
Our main result is that QED$_d$ in $d=2+\epsilon$ exhibits an intermediate phase \emph{between} the symmetric conformal large-$N$ phase and the chiral-symmetry-broken low-$N$ phase.
We thus distinguish between the conformal-critical flavor number $\Nconf$ below which the conformal phase becomes unstable and the chiral-critical flavor number $\Nchi$ below which chiral symmetry is broken.
In fact, the logical possibility that $\Nchi$ and $\Nconf$ could differ had been recognized earlier~\cite{braun2014}. It had, however, not been possible either to verify this eventuality, or to elaborate the nature of a potential intermediate phase.
In the present work, we show, within a controlled $\epsilon$ expansion around the \emph{lower} critical space-time dimension of $d=2$, that in fact $\Nconf > \Nchi$, at least when $\epsilon$ is small.
We demonstrate that the novel phase in between $\Nchi$ and $\Nconf$ is characterized by spontaneously broken Lorentz symmetry. Further independent arguments based on the $F$ theorem, mean-field theory, and perturbative expansion in fixed dimension, back up this conclusion, and lead to the prediction that the Lorentz-symmetry-breaking ground state exists also in the physical situation for $d=3$.
The resulting phase diagram in the $(d, N)$ plane is depicted in Fig.~\ref{fig:phase-diagram}.

\section{Model} 

Consider a U(1) gauge field coupled to $N$ flavors of massless Dirac fermions, described by the action
\begin{equation} \label{eq:action}
S_{\text{QED}} = \int d^dx \left(\frac{1}{4} F_{\mu\nu}F_{\mu\nu} + 
\bar\psi_i \gamma_\mu D_\mu \psi_i\right),
\end{equation}
where $i=1,\dots,N$ and $\mu,\nu = 0,\dots,d-1$, in $d$-dimensional Euclidean space-time with $2<d<4$. It is convenient to use four-component Dirac spinors $\psi$ and $\bar\psi \equiv \psi^\dagger \gamma_0$. The $4\times 4$ gamma matrices fulfill $\{\gamma_\mu,\gamma_\nu\} = 2 \delta_{\mu\nu} \mathbbm 1_{4}$. The field-strength tensor $F_{\mu\nu}$ involves the gauge field $A_\mu$ via the usual relation $F_{\mu\nu} = \partial_\mu A_\nu - \partial_\nu A_\mu$, and
the covariant derivative is $D_\mu = \partial_\mu + i e A_\mu$. $e$ is the electric charge. It has canonical mass dimension $[e^2] = 4-d$ and is therefore RG relevant towards the infrared. Put differently, QED$_3$ is asymptotically free, as QCD$_4$, but in contrast to QED$_4$. 
The theory enjoys a global ``chiral'' $\mathrm{SU}(2N)$ symmetry which basically is a consequence of the reducible representation of the Clifford algebra~\cite{gies2010}. The representation ensures that parity symmetry is preserved, and the Chern-Simons term is absent.
Furthermore, the theory respects the full relativistic invariance and the U(1) gauge symmetry. 
To fix the gauge, we add to the action $S_\text{gf} = - \frac{1}{2\xi} \int d^dx \left(\partial_\mu A_\mu \right)^2$, with undetermined $\xi$.

\section{RG flow} 

Once radiative corrections are taken into account, further interactions not present in the initial action may be generated from the RG flow. Most of them are RG irrelevant and can be neglected. Local four-fermion operators, however, are marginal in two dimensions. Within an $\epsilon$ expansion in $d=2+\epsilon$, they have to be taken into account, as they may play a decisive role at any interacting fixed point~\cite{footnote1}. The number of such interactions that are possible is strongly restricted by symmetry. In $d=3$, a full basis of fermionic four-point functions is given by~\cite{kubota2001, kaveh2005}
\begin{equation} \label{eq:four-fermi}
 S_\text{4-fermi} = \int d^dx \left[ g_1 (\bar\psi_i \gamma_{35} \psi_i)^2 + g_2 (\bar\psi_i \gamma_\mu \psi_i)^2 \right],
\end{equation}
where $\gamma_{35} \equiv i \gamma_3 \gamma_5$. $\gamma_3$ and $\gamma_5$ are the two ``left-over'' gamma matrices that have not been used to construct the fermion kinetic term. 
The operators present in Eq.~\eqref{eq:four-fermi} can be rewritten in terms of two-component spinors, allowing us to easily continue them to noninteger dimensions $2<d<4$~\cite{janssen2012, dipietro2016}.
Further operators are possible, but they can always be rewritten in terms of these by using Fierz identities~\cite{gies2010, braun2014}. We analyze the full theory space in which pure QED$_d$ represent one RG trajectory. By integrating the momentum shell from $\Lambda$ to $\Lambda/b$, we find the flow of the generalized action $S = S_\text{QED} + S_\text{gf} + S_\text{4-fermi}$ to one-loop order as
\begin{align} \label{eq:beta-e2}
 \frac{d e^2}{d \ln b} & = (4-d - \eta_A) e^2, \\
 \label{eq:beta-g1}
 \frac{d g_1}{d \ln b} & = (2-d) g_1 + \frac{16}{3} e^2 g_1 + 8 e^2 g_2 + 2 e^4 \nonumber\\
 & \quad - 4(2N-1) g_1^2 + 8 g_2^2 + 12 g_1 g_2, \\
 \label{eq:beta-g2}
 \frac{d g_2}{d \ln b} & = (2-d) g_2 + \frac{8}{3} e^2 g_1 + \frac{4}{3}(2N+1) g_2^2 + 4 g_1 g_2.
\end{align}
The gauge-field anomalous dimension is $\eta_A = \frac{4}{3} N e^2 + \mathcal O(e^4)$. In order to arrive at Eqs.~\eqref{eq:beta-e2}--\eqref{eq:beta-g2}, we have fixed the dimension of the Clifford-algebra representation as $\Tr \gamma_0^2 = 4$ and performed the angular integrations in $d=3$. The dimensions of the couplings are counted in general $d$~\cite{moon2013, janssen2015}. We have rescaled the couplings as $S_d / (2\pi)^d \Lambda^{d-4} e^2 \mapsto e^2$ and $S_d / (2\pi)^d \Lambda^{d-2} g_\alpha \mapsto g_\alpha$, with $S_d$ the surface area of the sphere in $d$ dimensions and $\Lambda$ the ultraviolet cutoff scale.
Importantly, we find that the beta functions are independent of the gauge-fixing parameter $\xi$, as they should be.
In the limit $N \to \infty$, Eqs.~\eqref{eq:beta-e2}--\eqref{eq:beta-g2} coincide with the previous large-$N$ result~\cite{kaveh2005}. At finite $N$, but $e^2 = 0$, Eqs.~\eqref{eq:beta-g1} and \eqref{eq:beta-g2} agree also with the flow equations of the purely fermionic system~\cite{gies2010}. Equation~\eqref{eq:beta-e2}, together with the equation for $\eta_A$, reduces to the universal one-loop flow of QED$_4$ in the limit $d\to 4$. In fact, we expect Eq.~\eqref{eq:beta-e2} to hold exactly as a consequence of the gauge invariance, in agreement with the analogous situation in the Abelian Higgs model~\cite{herbut1996}. 
At any interacting charged fixed point, we therefore must have $\eta_A^* = 4-d$ exactly, which fixes the fixed-point value of the charge. [At one loop, this leads to $e^2_* = \frac{3(4-d)}{4N} + \mathcal O(1/N^2)$.] The scaling form of the renormalized photon propagator, then, is $D_{\mu\nu}(q) \propto |q|^{2-d}$, which reproduces the finite photon mass of the Schwinger model in $d=2$~\cite{schwinger1962}, as well as the $1/|q|$ behavior in $d=3$~\cite{pisarski1984}.

\section{Conformal-critical flavor number $\Nconf$} 

We now evaluate Eqs.~\eqref{eq:beta-e2}--\eqref{eq:beta-g2} for $1/N \leq \epsilon \ll 1$, where $\epsilon = d-2$. In this limit, the fixed-point structure can be elucidated analytically by recognizing that
\begin{equation} \label{eq:fixed-point-e2-g1}
(e_*^2, g_1^*) = \left(\frac{3}{2N}, \frac{3}{4N^{3/2}}\right) + \mathcal O(\epsilon/N)
\end{equation}
defines a RG-invariant and infrared-attractive one-dimensional subspace of theory space. In fact, starting the RG with vanishing initial four-fermion couplings (pure QED$_d$), the couplings always flow towards this subspace. The remaining flow within the subspace is
\begin{equation} \label{eq:beta-g2-invariant-subspace}
 \left.\frac{d g_2}{d \ln b}\right|_{(e^2_*,g_1^*)} = -\epsilon g_2 + \frac{8N}{3} g_2^2 + \frac{3}{N^{5/2}}.
\end{equation}
There are therefore two fixed points
\begin{equation} \label{eq:fixed-point-g2}
 g_{2,\pm}^* = 
 \frac{3\epsilon}{16N}\left(1 \pm \sqrt{1-\left(\frac{\Nconf}{N}\right)^{3/2}} \right) + \mathcal O(\epsilon/N^{3/2})
\end{equation}
located at real couplings if 
\begin{equation}
 N\geq \Nconf(\epsilon) = \frac{8\sqrt[3]{2}}{\epsilon^{4/3}}.
\end{equation}
The fixed point at $(e_*^2, g_1^*, g_{2,-}^*)$ is entirely RG attractive and describes the conformal phase of QED$_d$. The phase has gapless fermionic and gauge-field excitations, just as the noninteracting system, but with nontrivial exponents, e.g., $\eta_A = 4-d$. 
By contrast, the second fixed point at $(e_*^2, g_1^*, g_{2,+}^*)$ has precisely one RG relevant direction, parallel to the $g_2$ axis. It describes a quantum critical point (QCP). For $N \searrow \Nconf$, the two fixed points merge and eventually disappear into the complex plane for $N < \Nconf$. If we now again run the RG for $N$ below but not too far from $\Nconf$, starting with the pure QED$_d$ action, the flow only slows down at $g_2 \approx \frac{3\epsilon}{16N}$, but ultimately runs towards divergent coupling $g_2 \to \infty$ at a finite RG scale, signaling the instability of the conformal state. 
This fixed-point annihilation scenario is not new: It has been proposed for various conformal gauge theories~\cite{gies2006, kaplan2009, kubota2001, braun2014}, often with application to condensed-matter systems~\cite{halperin1974, kaveh2005, herbut2014}. 
Note that we have found that $\Nconf$ diverges when $d\searrow 2$, in qualitative agreement with the expectation formulated in Ref.~\cite{dipietro2016}.
This is important, since the fixed-point annihilation now occurs in the perturbative regime, in which our expansion is under control.
In the following, we show that the phase below $\Nconf$ \emph{cannot} exhibit chiral symmetry breaking.

\section{Upper bound for chiral-critical flavor number $\Nchi$}

It has recently been demonstrated that the sphere free energy $F = - \log Z_{S^d}$ can be analytically continued to continuous dimension by defining $\tilde F = - \sin(\pi d/2) F$~\cite{giombi2015a}. ($Z_{S^d}$ is the partition function of the Euclidean conformal field theory on the $d$-dimensional sphere.) $\tilde F$ reduces to $\pi/2$ times the $a$-anomaly coefficient in $d=4$, and to $\pi/6$ times the central charge $c$ in $d=2$. $\tilde F = F$ in $d=3$. For integer dimension, therefore, formal proofs exist that show that $\tilde F$ decreases under the RG, $\tilde F_\mathrm{UV} > \tilde F_\mathrm{IR}$~\cite{zamolodchikov1986, myers2011, jafferis2011, casini2012, komargodski2011}. For continuous dimension, $\tilde F$ smoothly interpolates between the integer-$d$ limits. A body of evidence from various models supports the conjecture that monotonicity of $\tilde F$ holds also in general continuous $d$~\cite{giombi2015a, giombi2015b, fei2015, fei2016}. 
Assuming it in $d=2+\epsilon$ allows us to significantly constrain the possible values of $N$ at which chiral symmetry breaking can occur~\cite{appelquistcomment}. To leading order in $\epsilon$, we find for the conformal state~\cite{giombi2015b, giombi2016}
\begin{equation}
 \tilde F_\mathrm{conf} 
 = \left(N-\frac{1}{2}\right) \tilde F_\mathrm{f}
 = (2N-1)\frac{\pi}{6} + \mathcal O(\epsilon),
\end{equation}
where $\tilde F_\mathrm{f} = \pi/3 + \mathcal O(\epsilon)$ is the contribution of a single free massless Dirac fermion in the four-component representation~\cite{giombi2015a}. 
A state that spontaneously breaks chiral symmetry according to the pattern $\mathrm{U}(2N) \to \mathrm{U}(N) \times \mathrm{U}(N)$~\cite{appelquist1988, vafa1984, footnote4} would have $2N^2$ Goldstone modes and a massless transverse photon, contributing as~\cite{giombi2015b, giombi2016}
\begin{equation}
 \tilde F_\mathrm{\chi SB} 
 = \left(2N^2+(d-2)\right) \tilde F_\mathrm{b} 
 = N^2 \frac{\pi}{3} + \mathcal O(\epsilon).
\end{equation}
Here, $\tilde F_\mathrm{b} = \pi/6 + \mathcal O(\epsilon)$ is the contribution of a single free massless scalar boson, and the photon term can be understood as the contribution of $d-2$ independent scalar degrees of freedom~\cite{giombi2016}.
The noninteracting Gaussian theory accordingly has $\tilde F_\mathrm{UV} = N \tilde F_\mathrm{f} = N \frac{\pi}{3} + \mathcal O(\epsilon)$. To leading order in $\epsilon=d-2$, we thus have $\tilde F_\mathrm{UV} > \tilde F_\mathrm{conf}$ for all $N$, but $\tilde F_\mathrm{UV} > \tilde F_\mathrm{\chi SB}$ only for $N < 1$.
When $N > \Nconf$, the proposed flow from $\tilde F_\mathrm{UV}$ towards $\tilde F_\mathrm{conf}$ is thus consistent with the $F$ theorem, in agreement with the previous results~\cite{grover2014, giombi2015b, giombi2016}.
When $1<N<\Nconf$, however, with the conformal state being unstable, the $F$ theorem forbids a flow towards the chiral-symmetry-broken state. We thus find $\Nchi \leq 1 + \mathcal O(d-2)$, and consequently $\Nchi < \Nconf$, at least as long as $\epsilon$ is small.

\section{Intermediate phase}

According to the Vafa-Witten theorem~\cite{vafa1984}, QED$_d$ should have an unbroken $\mathrm{U}(N)\times \mathrm{U}(N)$ symmetry in the infrared and a gapless spectrum. This rules out the possibility that the fermions acquire a gap due to parity symmetry breaking, as well as other conceivable chiral breaking patterns with fewer than $2N^2$ Goldstone modes.
We now already see that the only possible symmetry-breaking pattern that is consistent with the $F$ theorem as well as the Vafa-Witten theorem is a spontaneous breakdown of Lorentz symmetry. In fact, it is precisely this conclusion that one is independently led to upon inspection of the flow of the four-fermion couplings:
From Eqs.~\eqref{eq:fixed-point-e2-g1} and \eqref{eq:beta-g2-invariant-subspace} we find that the flow for $N<\Nconf$ is always towards divergent $g_2$, but finite $g_1$ and $e^2$. On a mean-field level, QED$_d$ is hence effectively described by a Thirring-type interaction alone,
\begin{equation} \label{eq:thirring}
 S_\text{Thirring} = \int d^d x \left[ \bar\psi_i \gamma_\mu \partial_\mu \psi_i + g_2 \left(\bar\psi_i \gamma_\mu \psi_i\right)^2 \right],
\end{equation}
with positive $g_2 > 0$, corresponding to an \emph{attractive} interaction between like-charged fermions. The effect is similar to the situation in doped graphene, in which repulsive electron-electron interactions may induce short-range interactions in an attractive channel, eventually leading to, in this case, a superconducting instability~\cite{nandkishore2012}.
In the present system, this seemingly counterintuitive effect is evident already in the flow equation for $g_2$ with $e^2$ and $g_1$ set to their infrared fixed-point values. 
Finite $e^2$ leads to a positive contribution to $\left.(d g_2)/(d \ln b)\right|_{(e^2_*, g_1^*)}$; see the last term in Eq.~\eqref{eq:beta-g2-invariant-subspace}.

The present effective theory, Eq.~\eqref{eq:thirring}, can be solved at large $N$
by the saddle-point method, which yields the mean-field energy for the real order parameter $V_\mu = -i \sqrt{\frac{g_2}{2}}\langle \bar\psi_i \gamma_\mu \psi_i \rangle$ as
\begin{multline}
 \frac{f_\text{MF}(v)}{\Lambda^3} = \frac{1}{4g_2 \Lambda} v^2 - \frac{N S_3}{(2\pi)^3} \biggl[
 -1 + \frac{4}{3} v^2
 \\
 +\frac{1+2v^2}{2v} \ln \left(1+\frac{2v}{v^2-v+1}\right) 
 \\
  + \frac{2}{3}\ln\left(1+v^2+v^4\right) + \frac{v^3}{6} \artanh \frac{v}{1+v^2}
 \\
 + \frac{\sqrt{3}}{2} v^3 \left(\arctan\frac{v-2}{\sqrt{3} v} - \arctan\frac{v+2}{\sqrt{3}v}\right)
 \biggr],
\end{multline}
in $d=3$, where $v \equiv \sqrt{\frac{g_2}{2} V_\mu^2}/\Lambda$. Upon rescaling $S_3/(2\pi)^3 \Lambda g_2 \mapsto g_2$ and $f_\text{MF} \Lambda^{-3} (2\pi)^3/S_3 \mapsto f_\text{MF}$, and expanding in small $v$, we obtain
\begin{equation}
 f_\text{MF}(v) = \frac{1}{2} \left( \frac{1}{2g_2} - \frac{20 N}{3} \right) v^2 + \frac{\sqrt{3}}{2} \pi N |v|^3 + \mathcal O(v^4).
\end{equation}
%
%
At sufficiently large $g_2>0$, the mass of the vector order parameter $v$ thus changes sign, indicating a continuous transition towards a state with long-range order $i\langle \bar\psi_i \gamma_\mu \psi_i \rangle \neq 0$.
By contrast, the masses of all other possible order parameters remain positive.
The spectrum of this state has gapless fermions, but an anisotropic fermion propagator $G_\psi(q) = \frac{-i(\slashed{q}+\sqrt{2g_2}\slashed{V})}{(q_\mu + \sqrt{2g_2}V_\mu)^2}$, and thus the state is characterized by spontaneous Lorentz symmetry breaking.
The residual space-time rotation about the $V_\mu$ axis remains preserved. 
%
%
If, for instance, the order parameter has a temporal component, $V_0 \neq 0$, in a condensed-matter system such a state would reveal itself experimentally through the spontaneous formation of charge order. The situation may be similar to one of the various charge-modulated ground states reported in the fermionic honeycomb lattice model~\cite{motruk2015}, albeit without opening of a fermion gap.
Analogous Lorentz-symmetry-breaking condensates have previously also been found in effective theories in 3+1 dimensions~\cite{jenkins2004}. There, however, an attractive fermion-fermion interaction with $g_2 > 0$ had to be assumed from the outset. In the present system, by contrast, a positive coupling $g_2$ is generated by a finite repulsive gauge interaction $e^2$ during the RG.

\begin{figure}[tb]
\includegraphics[width=\linewidth]{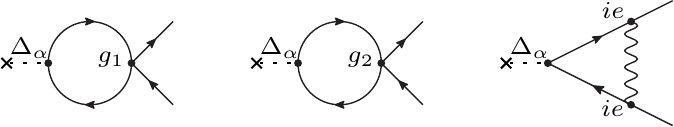}
\caption{Diagrams that contribute to the flow of symmmetry-breaking bilinears. Straight (wiggly) inner lines correspond to fermion (gauge) propagators. Dashed lines indicate bilinear insertions $\Delta_\alpha \in \{\Delta_{\mathrm{\chi SB}}, \Delta_{\mathrm{PSB}}, \Delta_{\mathrm{Kek}}, \Delta_{\mathrm{LSB}}^\mu\}$.}
\label{fig:susceptibilities}
\end{figure}

Within the $2+\epsilon$ expansion, we can in fact demonstrate the emergence of the Lorentz-symmetry-breaking state in a controlled way by a susceptibility analysis~\cite{furukawa1998, salmhofer2004, vafek2010, nandkishore2012}.
To this end, we add to the action, Eq.~\eqref{eq:action}, infinitesimally small symmetry-breaking bilinears
\begin{align}
 S_\Delta & = \int d^dx\, \bar\psi_i \bigl[ 
 \Delta_\mathrm{\chi SB} \mathbbm 1_4
 + \Delta_\mathrm{PSB} \gamma_{35} 
 \nonumber \\ & \quad
 + i \Delta_{\mathrm{Kek}} \left(\gamma_3\cos \varphi  + \gamma_5 \sin\varphi \right)
 + i \Delta^\mu_\mathrm{LSB} \gamma_\mu
 \bigr]  \psi_i.
\end{align}
Finite $\Delta_\mathrm{\chi SB}$ breaks chiral symmetry, $\Delta_\mathrm{PSB}$ breaks parity and time reversal, the ``Kekul\'e''~\cite{hou2007} mass $\Delta_\mathrm{Kek}$ also breaks chiral symmetry (but a different subset), and $\Delta_\mathrm{LSB}$ breaks Lorentz symmetry.
Evaluating the diagrams in Fig.~\ref{fig:susceptibilities}, we find the leading-order flow equations~\cite{footnote1b}
\begin{figure*}[t]
 \begin{center}
 \includegraphics[width=\textwidth]{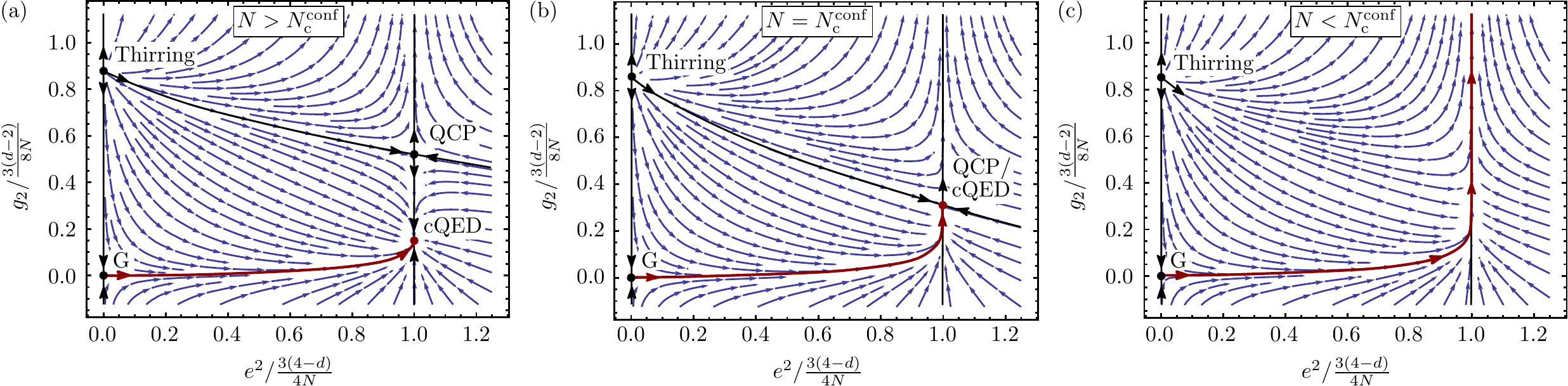}
 \end{center}
 \caption{RG flow in $d=3$ for (a) $N = 7$, (b) $N = 6.24$, and (c) $N = 6$. Arrows point towards infrared. The RG trajectory of pure QED$_3$ is depicted as a red line. Here, we have chosen $g_1$ such that $(d g_1) / (d \ln b) = 0$. At large $N > \Nconf$, there are three interacting fixed points besides the Gaussian (G): a purely fermionic fixed point (Thirring) at positive $g_2^*$, which can be understood as the ultraviolet completion of the three-dimensional Thirring model~\cite{gies2010}, the infrared attractive conformal fixed point (cQED) at finite $e_*^2$, and a quantum critical point (QCP) with one RG relevant direction (a). Upon lowering $N$, the QCP and cQED fixed points approach each other and eventually merge at $N = \Nconf$ (b). In the one-loop approximation, this happens at $\Nconf(3) \approx 6.24$. For $N < \Nconf$, the fixed points annihilate and disappear into the complex plane, leaving behind the runaway flow (c).}
 \label{fig:rg-flow}
\end{figure*}
\begin{align}
 \frac{d \Delta_\mathrm{\chi SB}}{d \ln b} & = \left(1 + 2 g_1 +6 g_2 + \frac{8}{3} e^2 \right) \Delta_\mathrm{\chi SB}, \label{eq:beta-Delta-chiSB}\\
 \frac{d \Delta_\mathrm{PSB}}{d \ln b} & = \left(1 - 2(4N-1) g_1 + 6 g_2 + \frac{8}{3} e^2 \right) \Delta_\mathrm{PSB}, \\
 \frac{d \Delta_\mathrm{Kek}}{d \ln b} & = \left(1 - 2 g_1 + 6 g_2 + \frac{8}{3} e^2 \right) \Delta_\mathrm{Kek}, \\
 \frac{d \Delta_\mathrm{LSB}^\mu}{d \ln b} & = \left(1 - \frac{2}{3} g_1 + \frac{2}{3}(4N+1) g_2 \right) \Delta_\mathrm{LSB}^\mu. \label{eq:beta-Delta-LSB}
\end{align}
Now consider $N$ near and below $\Nconf$ and infinitesimally small $\Delta$'s. In this limit, the flow approaches the fixed-point regime at $(e^2_*, g_1^*, \text{Re}(g_{2,\pm}^*))$, slows down, and eventually runs off towards infinite $g_2$. The infrared behavior then does not depend on whether we start with initial $g_1 = g_2 = 0$ (pure QED$_d$) for $N \lesssim \Nconf$ or at strong coupling above the fixed point for $N \gtrsim \Nconf$.
We therefore determine the dominant ordering by computing the largest susceptibility \emph{at the QCP} located at $(e^2_*, g_1^*, g_{2,+}^*)$, for $N \searrow \Nconf$.
The scaling form of the free energy is 
\begin{equation}
f(\delta \vec g, \Delta_\alpha) = \lvert \delta \vec g \rvert^{d/y} \,\Phi_\alpha^\pm\! \left( \frac{\Delta_\alpha}{\lvert \delta \vec g \rvert^{x_\alpha/y}}\right)
\end{equation}
with the scaling function $\Phi^\pm_\alpha$, and
where $\delta \vec g = (\delta e^2, \delta g_1, \delta g_2)$ is the eigenvector associated with the unique RG relevant direction at the QCP and $y$ its corresponding eigenvalue, $\frac{d \delta \vec g}{d \ln b} = y \, \delta \vec g + \mathcal O(\delta g^2)$. $x_\alpha$ is the eigenvalue associated with the RG flow of $\Delta_\alpha$, $\frac{d \Delta_\alpha}{d\ln b} = x_\alpha \Delta_\alpha$.
The scaling of the susceptibility $\chi$ is therefore
\begin{equation}
 \chi_\alpha = - \frac{\partial^2 f}{\partial \Delta_\alpha^2} \propto |\delta \vec g_0|^{-\gamma_\alpha}, \quad \text{with } \quad \gamma_\alpha = \frac{2x_\alpha - d}{y},
\end{equation}
where $|\delta \vec g_0| = |\vec g - \vec g_\mathrm{c}|$ is the distance to criticality.
At the QCP [Eqs.~\eqref{eq:fixed-point-e2-g1} and \eqref{eq:fixed-point-g2}] we find to the leading order in $\epsilon$ and for $N \geq \Nconf$:
\begin{align}
 \gamma_\mathrm{\chi SB}/\nu & = - \epsilon + \frac{8}{N} 
 + \mathcal O(\epsilon^{3/2}) = \gamma_\mathrm{Kek}/\nu, \\
 \gamma_\mathrm{PSB}/\nu & = - \epsilon - \frac{12}{\sqrt{N}} + \frac{8}{N}
 + \mathcal O(\epsilon^{3/2}), \\
 \gamma_\mathrm{LSB}/\nu & = \epsilon \sqrt{1 - \left(\frac{\Nconf}{N}\right)^{3/2}}
 + \mathcal O(\epsilon^2), \label{eq:gamma-LSB}
\end{align}
where $\nu = 1/y = 1/(\epsilon \sqrt{1-(\Nconf/N)^{3/2}})$ is the leading-order correlation-length exponent, which diverges upon $N\searrow N_\mathrm{c}^\text{conf}$~\cite{herbut2016}. Superconducting gaps may be considered as well, but also lead to negative $\gamma$. For all $N > \Nconf(\epsilon)$, the only bilinear that leads to a non-negative exponent at the QCP is the one that breaks Lorentz symmetry, with $\gamma_\mathrm{LSB} = 1$ at leading order in $\epsilon$.

\section{Towards $d=3$}

Within a perturbative approach, we may evaluate Eqs.~\eqref{eq:beta-e2}--\eqref{eq:beta-g2} also directly in fixed $d\in(2,4)$. Again, we find that the conformal fixed point exists only above a certain $\Nconf(d)$, at which it merges with the quantum critical point. In $d=3$ this happens at $\Nconf(3) \approx 6.24$. 
The RG flow projected onto the $e^2$-$g_2$ plane is depicted for different values of $N$ near $\Nconf$ in Fig.~\ref{fig:rg-flow}.
In $d<3$, the perturbative result for $\Nconf(d)$ approaches the value from the $2+\epsilon$ expansion; see Fig.~\ref{fig:phase-diagram}. In $d\nearrow 4$, we find $\Nconf \propto 4-d$, in agreement with the recent results from $4-\epsilon$ expansion~\cite{dipietro2016, herbut2016}.

For $N$ near and below $\Nconf$, the RG flow approaches the regime of the (then complex) fixed points, slows down, and eventually runs off towards large four-fermion coupling, precisely as in $d=2+\epsilon$. How low does $\Nconf$ have to be such that a flow from the Gaussian fixed point through the (conformal) fixed-point regime towards a chiral-symmetry-broken infrared regime would be allowed? The $F$ theorem gives an upper bound: Directly in $d=3$, we have 
$F_\mathrm{conf} = N F_\mathrm{f} + \frac{1}{2} \log \left(\frac{\pi N}{4}\right) + \mathcal O(N^{-1})$,
where $F_\mathrm{f} = \frac{\log 2}{2} + \frac{3\zeta(3)}{4\pi^2}$~\cite{klebanov2012, giombi2015b}. The chiral-symmetry-broken state has $F_\mathrm{\chi SB} = (2N^2 + 1) F_\mathrm{b}$, with $F_\mathrm{b} = \frac{\log 2}{8} - \frac{3\zeta(3)}{16\pi^2}$. From the requirement $F_\mathrm{conf} > F_\mathrm{\chi SB}$, we find $\Nchi \leq 4.422$~\cite{footnote2}. This is in $<1\permil$ agreement with the value obtained from the (Pad\'e resummed) $4-\epsilon$ expansion of $\tilde F_\mathrm{conf}$~\cite{giombi2015b}. We therefore expect this estimate for an upper bound on $\Nchi$ in $d=3$ to be quite accurate, the neglected higher-order terms in $F_\mathrm{conf}$ notwithstanding. (For continuous $d$, we can similarly find an upper bound for $\Nchi$ from the analytic continuation of the $F$ theorem~\cite{giombi2015b}; see Fig.~\ref{fig:phase-diagram}.)
Thus, also in the physical situation of $d=3$, there appears to be a finite range of values of $N$ in which the only possible ground state consistent with the $F$ and Vafa-Witten theorems, as well as the instability of the conformal state, is the Lorentz-symmetry-breaking state.

\section{Conclusions}

We have argued that quantum electrodynamics in three space-time dimensions exhibits a previously overlooked phase at intermediate flavor number. This phase is located between the conformal phase at large $N>\Nconf$ and the chiral-symmetry-broken phase at small $N<\Nchi$. 
This is the agreeing result of four independent approaches to the problem: 
(i) $2+\epsilon$ expansion shows that the conformal phase is unstable below a certain $\Nconf$. The susceptibility analysis proves that the instability is towards the Lorentz-symmetry-breaking state. This result is under control as long as $\epsilon$ is small.
(ii) When the conformal state is unstable, the only phase that is compatible with both the $F$ theorem and the Vafa-Witten theorem is the Lorentz-symmetry-breaking state.  In particular, chiral symmetry breaking and plain parity symmetry breaking are forbidden. In $d=2+\epsilon$, this is a rigorous statement assuming that the generalized $F$ theorem, which has formally been proven only for integer dimension, holds also in its dimensionally continued version.
(iii) Spontaneous Lorentz symmetry breaking is also predicted by mean-field theory in $d=3$. This is controlled for large $N$, in which order-parameter fluctuations are suppressed.
(iv) Finally, indications for the emergence of the Lorentz-symmetry-breaking phase are also obtained by a simple perturbative expansion in fixed $d=3$.
We note that our result is in qualitative agreement with the most recent numerical results, which find an instability of the conformal state~\cite{raviv2014}, but no evidence for a chiral-symmetry-breaking condensate at low values of~$N$~\cite{karthik2016, footnote3}.

Our findings have significant implications for the condensed-matter systems that are described by QED$_3$: U(1) spin liquids, for instance, are much more prone to destabilize than previously thought~\cite{grover2014}. Surface states of clean topological insulators might display unusual interaction effects, such as spontaneous formation of electron-hole puddles or emergent anisotropy.

Spontaneous breaking of Lorentz symmetry has intriguing consequences when the symmetry becomes local (as in a theory with gravity)~\cite{bluhm2005}. A natural question that arises is the fate of the Goldstone modes: Can a Higgs mechanism occur? Can a gauge field (``photon'') emerge as a Goldstone mode associated with the spontaneous Lorentz symmetry breaking~\cite{bjorken1963, bjorken2001}? QED$_3$ might provide a new playground to study such problems.

\begin{acknowledgments}
I am grateful to Subhro Bhattacharjee, Shailesh Chandrasekharan, Holger Gies, Simon Hands, Yin-Chen He, Igor Herbut, Timo L\"ahde, Achim Rosch, Dietrich Roscher, Matthias Vojta, and Andreas Wipf for very helpful discussions, and to Holger Gies and Matthias Vojta for insightful comments on the manuscript. I also profited from a collaboration with Jens Braun, Holger Gies, and Dietrich Roscher on an earlier related project. This work was supported by the DFG under SFB\,1143 ``Correlated Magnetism: From Frustration to Topology.''
\end{acknowledgments}


\begin{thebibliography}{99}
\bibitem{footnote1}
It has recently been demonstrated that even in $d=4-\epsilon$ it is precisely these quartic operators which drive the instability of the conformal phase~\cite{dipietro2016}.
\bibitem{footnote4}
Here, $\mathrm{U}(2N) = \mathrm{SU}(2N) \times \mathrm{U}(1)$ where $\mathrm U(1)$ is the global part of the gauge symmetry.
\bibitem{footnote2}
Note that the quoted value restricts only the flow from the regime of the conformal fixed point towards the chirally broken state; i.e., it is valid in the vicinity of $\Nconf$.
\bibitem{footnote3}
We note, however, that after this work was posted on the arXiv a preprint appeared claiming that $\Nconf < 1$ in $d=3$~\cite{karthik2016b}, in disagreement with Ref.~\cite{raviv2014}.
\end{thebibliography}
\end{document}